\documentclass[12pt]{iopart}

\usepackage{iopams,color,ulem}

\usepackage{graphics}
\usepackage{graphicx}
\usepackage{epsfig,color}

\definecolor{pdcolor}{rgb}{1,0.5,0}

\definecolor{pdblue}{rgb}{0,0,1}

\definecolor{rkgreen}{rgb}{0,1,0}

\begin{document}

\title[CLT: higher-order anomalies]{In the folds of  the Central Limit Theorem:\\
  L\'evy walks, large deviations and higher-order anomalous diffusion}
\author{Massimiliano Giona$^1$, Andrea Cairoli$^2$ and Rainer Klages$^{3*}$}
\address{$^1$Dipartimento di Ingegneria Chimica, Materiali, Ambiente,
La Sapienza Universit\`a di Roma, Via Eudossiana 18, 00184 Roma, Italy\\
 E-mail: massimiliano.giona@uniroma1.it\\
 $^2$ The Francis Crick Institute, 1 Midland Road, London NW1 1AT, United Kingdom\\
 E-mail:  andrea.cairoli@crick.ac.uk\\
$^3$School of Mathematical Sciences, Queen Mary University of London,
Mile End Road, London E1 4NS, United Kingdom\\
$^*$ corresponding author}
\ead{r.klages@qmul.ac.uk}
\vspace{10pt}
\begin{indented}
\item[]
\end{indented}

\begin{abstract}
This article considers the statistical properties
of L\'evy walks possessing a regular long-term linear scaling of the
mean square displacement with time, for which the conditions of the classical
Central Limit Theorem apply. Notwithstanding this property,
their higher-order moments display anomalous scaling
properties, whenever the statistics of the transition times possesses
power-law tails. 
This phenomenon is  perfectly consistent with the classical Central Limit Theorem,
as it involves the convergence properties towards the normal distribution. 
This phenomenon is closely related to the property  that the higher order moments  
of normalized sums of $N$ independent random variables possessing
finite variance may deviate, for $N$ tending to infinity,
 to those of the 
normal distribution.  
 The thermodynamic implications of these results are
thoroughly analyzed by motivating the concept of higher-order
anomalous diffusion.
\end{abstract}

\section{Introduction}
\label{sec1}
The Central Limit Theorem (CLT) constitutes 
a highly articulated and subtle galaxy of results that originated from a simple 
core concept, representing a milestone  of the theory of
probability and statistical physics because of its general and
widespread applicability.  In its classical 
form, formulated for independent and identically distributed (iid) random variables,
the basic principle can be 
stated  as follows
  \cite{clt1,clt2}: Consider iid random variables
  $\{x_h\}_{h=1,\ldots,N}$,  with
  zero mean and finite variance, and their normalized sum,
\begin{equation}
z= \frac{1}{a(N)} \sum_{h=1}^N x_h\quad ,
\label{eq1}
\end{equation}
where $a(N)= \sqrt{N} \, \langle x^2 \rangle$ and $\langle \cdot
\rangle$ denotes the expected value with respect to the probability
density of the variables $x_h$.  Then the probability density
function of $z$, $p(z;N)$, converges in distribution to the normal distribution
$g(z)=e^{-z^2/2}/\sqrt{2 \, \pi}$ for $ N \rightarrow \infty$.  We
refer the interested reader to \cite{clt3} for a chronological
overview of various approaches to derive limit theorems and
convergence properties of sums of random variables.  The CLT and the
occurrence of normal distributions yield a paradigmatic crossroad
between mathematics and physics, as formulated by M.~Kac (quoting
H.~Poincar\'e), ``...there must be something mysterious about the
normal law, since mathematicians think it is a law of nature whereas
physicists are convinced that it is a mathematical theorem''
\cite{kaciid}. 

In this form,  the CLT represents the probabilistic interpretation
of the macroscopic properties of Brownian motion,  pioneered by Einstein and Smoluchowski,
providing   the connection between Brownian motion and  diffusive phenomena \cite{einstein,smoluch}.
If the conditions of the CLT are not met, new physics appears in connection
with molecular and particle motion, usually referred to as {\em anomalous diffusion}.
This is the case  where the
variance of the variables $x_h$ is unbounded \cite{unbound}.  In this
context, limit distributions different from the Gaussian typically
describe the position density functions of random walks with power law
tailed distributed hopping lengths and/or transition times between two
consecutive changes in the direction of motion
\cite{ctrw,anomalous,anomalous1}. This phenomenon can be mathematically
interpreted
by the Generalized CLT \cite{clt1}, involving
$\alpha$-stable distributions  as limit distributions.
Other cases    of deviation from normality involve the random sums  of
independent random variables \cite{rans1,rans2}, but this case is of
limited interest in the present analysis.

In this article we consider the statistical properties
  of random motions, possessing bounded propagation velocity and
  linear asymptotic scaling of the mean square displacement.  Despite
  the fact that these processes are usually classified as {\em regular
    diffusion}, we show that anomalies in their statistical properties
  arise when higher-order moments are considered.  These anomalies are
  in one-to-one relation to the convergence properties of the density
  function of sums of independent random variable towards the normal
  distribution.  In this respect, we propose a new class to which
  these processes should be referred to.

Natural candidates for this analysis are represented by L\'evy Walks
(LWs) \cite{lw1,lw2}.  By definition, these processes possess a
bounded propagation velocity and are characterized by a prescribed
probability density of the transition times $T(\tau)$ \cite{lw3}.  In
the literature, the focus has been so far almost exclusively on the
case where the second-order moment of $T(\tau)$ is unbounded.  This in
fact provides a classical example of violation of the assumptions
underlying the CLT, thus leading to anomalous diffusion
\cite{lw2,schmiedeberg,lw3,lw4,lw_barkai,fouxon17,vezzani19,miron20}.
For the resulting processes in this case, the concept of an infinite
density has been introduced to describe the long-term properties of
the probability density function \cite{lw4,lw_barkai}, and interesting
connections with the deterministic Lorentz gas model have been
explored \cite{zarfaty18,fouxon20}. Alternatively, these properties
can also be obtained from the big jump principle, capturing rare
events beyond traditional CLTs \cite{vezzani19}.

However, LWs exhibiting a long-term scaling of the mean square displacement that is linear in time,
which we call Einsteinian, have been rather overlooked.
They have been discussed in only few articles (e.g., \cite{rebenshtok16}),
despite the fact that they are characterized by peculiar features as regards
the scaling of the generalized moments. 
These depend generically on the
convergence properties of the distribution tails.
Reference~\cite{wang20} studies this case in connection with a two-state
model of a random walk consisting of the stochastic superposition of a
LW and a Brownian motion process.

The main focus of this article is to investigate the scaling properties
of the moment hierarchy of Einsteinian LWs.
Our first main result is to show that these processes
exhibit highly anomalous features in the higher-order elements of the moment hierarchy,
while still preserving the long-term linear scaling of the mean square displacement and
without violating the basic conditions ensuring the application of the CLT.
We discuss how this phenomenon lies in the folds of the CLT,
and is associated with the large-deviation properties of the limiting
probability density function
in the long-time limit. 
To describe this phenomenology systematically,
we introduce the concept of {\it higher-order anomalous diffusion},
which is our second main result. In detail,
any LW for which the conditions of the CLT apply,
but whose transition time distribution 
possesses long-term power-law tails,
$T(\tau) \sim \tau^{-(\xi+1)}$ with $\xi>2$,
falls in this class.

The phenomenon of higher-order anomalous diffusion 
involves anomalies in the spectrum of generalised moments
of a random walk process $X(t)$.  Given a realization $x(t)$ of this
process, the generalised moments are defined as \cite{vulpio}
\begin{equation}
\langle |x(t)|^q \rangle \sim t^{q \, \nu(q)} \, ,
\qquad q \in {\mathbb R}^+\quad .
\label{eq1_1}
\end{equation}
Based on the structure of $\nu(q)$, we can distinguish three classes
of stochastic motions \cite{vulpio}: (i) $\nu(q) = 1/2$, which is
characteristic of regular diffusive processes.  (ii) $\nu(q) = \nu_0$,
where $\nu_0 \neq 1/2$ is a constant parameter independent of the
order $q$ of the generalized moment.  This property characterizes the
class of ``weak anomalous diffusive" processes.  (iii) $\nu(q)$
non-uniform, i.e., generically dependent on the exponent $q$.  This
property characterizes the family of ``strong anomalous diffusive"
processes.  Strong anomalous diffusion has been found in the context
of transport phenomena driven either by stochastic perturbations or by
deterministic chaotic forcings \cite{str0,str1,str2,str3,str4,voll}.  In
full generality, systems exibiting strong anomalous diffusion
properties display an almost discontinuous transition in the scaling
behaviour of the exponents of their generalized moments. Namely, they
satisfy the condition
\begin{equation}
q \, \nu(q) \sim
\left \{
\begin{array}{lll}
\nu_1 q  & \; \; \;  &  \mbox{for} \; q < q_c \\
\nu_2 q - k & \; \; \; & \mbox{for} \; q > q_c
\end{array}
\right . \quad ,
\label{eq1_2}
\end{equation}
with $\nu(2) \neq 1/2$, $\nu_2=1$ and constant $k$.  However, 
to the best of our knowledge,
we do not know of a diffusive dynamics for which $\nu(q)$ is a continuous
and smooth function of the power law index $q$.

The occurrence of higher-order anomalies has also deep
implications in the thermodynamic description of the process,
and this is one of the major physical issues involved in this
phenomenon. Despite the linear Einsteinian scaling, a correct hydrodynamic
limit for these processes, capable of reproducing the  temporal behavior of
the whole moment hierarchy  cannot be expressed simply by the diffusion equation.

The article is organized as follows.  In Section \ref{sec2} we
characterize the probability density function of sums of  $N$ iid
variables possessing bounded variance but unbounded
higher-order moments.  This is achieved by connecting this problem to
the large deviation properties of Einsteinian LWs.  
Specifically,  the  convergence properties of CLT  even in the simplest case of iid
random variables,
involve a singular limit, for $N$ tending to infinity, for the expected values
of unbounded functions (such as the monomials of  $z$). 
Section \ref{sec2bis} analyzes further this singular limit, by means of simple
and anaytically tractable examples.
Section \ref{sec3new} develops the  statistical description of the
moments  starting from the hyperbolic transport equations for LW.
Section \ref{sec3}
discusses the thermodynamic and transport implications of this result,
by constructing an approximate statistical two-phase model that
recovers the same scaling properties of the moment hierarchy of these
LWs.  We also introduce the concept of ``differential moment exponent
spectrum", a tool alternative to the spectrum $\nu(q)$ defined by
eq.~(\ref{eq1_1}), which provides a convenient way to highlight the
propagation mechanisms underlying anomalous diffusion properties of
LWs.  We conclude in Section~\ref{sec5} by summarising our results and
elaborating on the future perspectives.

\section{Large deviation properties of regular LWs}
\label{sec2} 

Adopting the formulation typical of Continuous Time Random Walks, 
a LW can be described as a process characterized by a single velocity, $v_0$, 
attaining at each transition instant either a positive or a
negative velocity direction with equal probability.  The statistics
of the transition time $\tau$, i.e., of the time interval between two
subsequent transitions, is specified by a prescribed probability density
function $T(\tau)$.  If $n \in {\mathbb N}$ indicates the operational
time, counting the number of transitions, and $x_n$, $t_n$ are the
walker position and the physical time after $n$ transitions,
respectively, the LW is described by the iterative stochastic
algorithm
\begin{equation}
x_{n}=x_{n-1} +  v_0 \, r_{n} \, \tau_{n} \, , \qquad
t_{n}=t_{n-1}+\tau_{n}
\label{eq2_1}
\end{equation}
where
 $\{ r_n \}_{n \in {\mathbb N}}$ is a family of iid random
variables attaining values $\pm 1$ with equal probabilities, and
$\{\tau_n\}_{n \in {\mathbb N}}$ is a sequence of iid random variables 
sampled from the distribution $T(\tau)$.
The two sequences are independent of one another. 
As initial condition at time $t_0=0$, we set $x_0=0$.
Without loss of generality, we also set $v_0=1$ a.u.

In a continuous time setting, considering the physical time $t \in
{\mathbb R}^+$, and assuming inertial particle motion (i.e., straight
line motion between two consecutive space-time points $(x_n,t_n)$,
$(x_{n+1},t_{n+1})$), the particle position $x(t)$ at time $t$ can be
expressed as
\begin{equation}
x(t)=  \sum_{h=1}^{N(t)}  \,  r_h \, \tau_h +
 \, r_{N(t)}  \chi_{N(t)}(t)
\label{eq2}
\end{equation}
where $\chi_{N(t)}= t- \sum_{h=1}^{N(t)} \tau_h $, and 
the integer $N(t)$ 
is defined by the inequality
$\sum_{h=1}^{N(t)} \tau_h <t \leq \sum_{h=1}^{N(t)+1} \tau_h$.

A widely used expression for a transition-time probability density
function possessing power-law tails is \cite{lw2}
\begin{equation}
T(\tau) = \frac{\xi}{\tau_0 \, (1 + \tau/\tau_0)^{\xi+1}}\quad ,
\label{eq3}
\end{equation}
with constant $\xi>0$ and 
$\tau_0=1$ a.u.
For this class of LWs, anomalous transport properties occur for $\xi
\in (0,2)$ \cite{lw3}. Indicating with $\sigma_x^2(t)= \langle x^2(t)
\rangle$ the mean square displacement (as from eq.~(\ref{eq2}),
$\langle x(t) \rangle =0$), one observes the following phenomenology:
for $\xi \in (0,1)$ a ballistic scaling, $\sigma_x^2(t) \sim t^2$; 
for $\xi \in (1,2)$ superdiffusive anomalous diffusion, $\sigma_x^2(t)= \sim t^{3-\xi}$;
for $\xi >2$ a regular linear scaling, $\sigma_x^2(t) \sim t$. 
We refer to the latter LW as {\it Einsteinian}, or regular.

For $\xi>2$ the CLT applies, so that the probability density function
for the normalized variable $x/\sigma_x(t)$ converges  in a weak sense to a normal
distribution \cite{billing}.  
This seemingly trivial phenomenology is the reason why this case 
has been scarcely considered in the literature so far.
In fact, even for these Einsteinian LWs, 
interesting and highly non trivial properties occur, owing to the power-law decay of the
transition-time probability density function $T(\tau)$.  
More precisely, we can show the existence of 
anomalous deviations in the scaling of the higher-order moments for any value of
$\xi>2$, without violating the assumptions of the CLT. This phenomenon has been shortly
addressed in \cite{rebenshtok16,wang20}, but its implications have not
been developed in detail.

We now formulate the mathematical setting for this problem, 
by considering the stochastic variable defined in eq.~(\ref{eq1}).  
The CLT states that the probability density function $p(z;N)$ associated
with $z$, (possessing zero mean and unit variance), defined as the  rescaled 
superposition of $N$ independent random
variables $x_h$, converges to the normalized Gaussian density
$g(z)=e^{-z^2/2}/\sqrt{2 \, \pi}$ in a distributional sense
\cite{clt1}. Given any smooth bounded test function
$\phi(z)$, this can be written as
\begin{equation}
\lim_{N \rightarrow \infty} \int_{-\infty}^\infty  \phi(z)  \left    
[p(z;N)- g(z)
\right ] \, d z =0
\label{eq4}
\end{equation}
If the test function is not bounded, e.g. $\phi(z)=z^{2 n}$
with $n >1$, we can still obtain 
\begin{equation}
\lim_{N \rightarrow \infty} \int_{-\infty}^\infty z^{2 \, n} \left [
p(z;N)-g(z) \right ] \, dz \neq 0
\label{eq5a}
\end{equation}
without violating the CLT. 
Eq. (\ref{eq5a}), that can be rewritten as,
\begin{equation}
\lim_{N \rightarrow \infty} \int_{-\infty}^\infty  z^{2 \, n}
\, p(z;N) \, d z \neq \int_{-\infty}^\infty z^{2 \, n} \, \left (
\lim_{N \rightarrow \infty} p(z;N) \right ) \, dz
\label{eq5b}
\end{equation}
implies that the limit for the number $N$ of the summed random variables
tending to infinity does not commute, in general, with the operation of taking the expected 
value  of unbounded functions of $z$ 
with respect to the  corresponding probability measure.
This corresponds to the fact that the
higher-order moments of $p(z;N)$ may deviate, for large $N$, from
those of a normal distribution.  
We refer to this phenomenon as 
an {\it anomalous moment scaling} 
or {\it higher-order anomalous diffusion}.  
We now demonstrate that this is a generic feature of LWs ultimately
associated with the transition time statistics defined by
eq.~(\ref{eq3}).

The number $N$ of iid random variables corresponds in
a CTRW-description to the  operational time of a LW; 
in a continuous time setting to the physical time $t$.
As such 
the behavior of
the moment hierarchy of a LW as a function of time $t$ enables us to
monitor indirectly the convergence properties of $p(z,t)$ where 
$z=x(t)/\langle x^2(t) \rangle$, 
and ultimately the
singular limit  defined by eq.  (\ref{eq5b}) underlying the  statistical
properties of the sum $\sum_{h=1}^N r_h \, \tau_h$ for finite $N$.
To show the existence of these anomalies, therefore, 
our strategy is summarised as follows: 
(i) First, we discuss the properties of $p(z;N)$ for finite $N$. 
(ii) Second, we take the moments of $z$ with respect to
$p(z;N)$ 
and analyze their scaling behavior as $N \rightarrow \infty$. 
(iii) Finally, we compare the result obtained with the r.h.s. of eq. (\ref{eq5b}), 
which are nothing but the moments of the standard normal distribution.

\subsection{Scaling analysis of the finite-$N$ distribution function $p(z;N)$.}
\label{sec2_0}

Consider first the case of the  sum of identical
independent random variables $\tau_h$, 
\begin{equation}
\eta_N = \sum_{h=1}^N {r_h} \tau_h\quad ,
\label{eq5_0}
\end{equation}
where $\tau_h$ are independent random variables distributed according
to eq.~(\ref{eq3}). Consider the normalized quantity
\begin{equation}
z_N=\frac{\eta_N-\langle \eta_N \rangle}{\sqrt{\langle \eta_N^2 \rangle - \langle \eta_N \rangle^{2}}}
\label{eq5}
\end{equation}
and let the exponent $\xi$ be greater than
$2$. Specifically, for $\xi \in (2,3)$ the third-order moment of
$T(\tau)$ is unbounded, but the second-order moment of
$\eta_N$ is still a linear function of $N$.

\begin{figure}[htbp]
\begin{center}
\resizebox{0.8\columnwidth}{!}{%
 \includegraphics{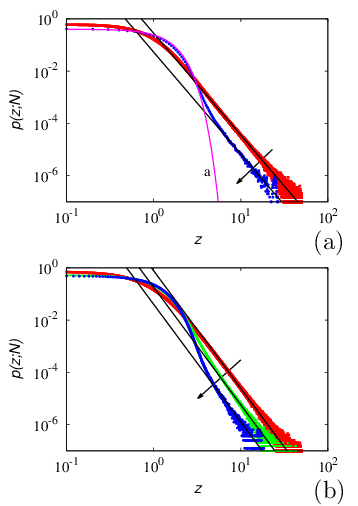}}
\end{center}
\caption{Probability density $p(z;N)$ vs.\ $z$ for the random variable
  $z_N$ defined by eq.~(\ref{eq5}) with $T(\tau)$ given by
  eq.~(\ref{eq3}). The arrows indicate increasing values of $N$.
Solid lines represent the  asymptotic scaling
  $P(z;N) \sim 1/z^{1+\xi}$.
Panel (a): $\xi=2.9$, $N=300,\, 3000$. Line (a) corresponds to the normalized
  Gaussian distribution $g(z)$.
 Panel (b):
  $\xi=3.5$, $N=10,\,100,\,1000$. 
We simulated $2 \times 10^8$ realizations of the process in all cases, 
  apart from the case $\xi=2.9$, $N =3000$ where we simulated $2 \times 10^7$ realisations.}
\label{Fig1}
\end{figure}

For sufficiently large $N$, the behavior of the density function
$p(z;N)$ for $z_N$  can be approximated by
\begin{equation}
p(z;N) \simeq
\left \{
\begin{array}{lll}
g(z) & \;\;\; & \mbox{for} \;  |z| < z_c(N) \\
 & \;\;\; & \\
\frac{C(N)}{z^{\xi+1}} & & \mbox{for} \; |z| > z_c(N)
\end{array}
\right .
\label{eq6}
\end{equation}
where $z_c(N) \rightarrow \infty$ and $C(N) \rightarrow 0$ for $N
\rightarrow \infty$, owing to the application of the CLT.  This
phenomenon is validated numerically in Fig.~\ref{Fig1}
 for two exemplary values of $\xi$: panel (a), $\xi=2.9$;
and panel (b), $\xi=3.5$.  
 
\begin{figure}[htbp]
\begin{center}
\resizebox{0.8\columnwidth}{!}{%
 \includegraphics{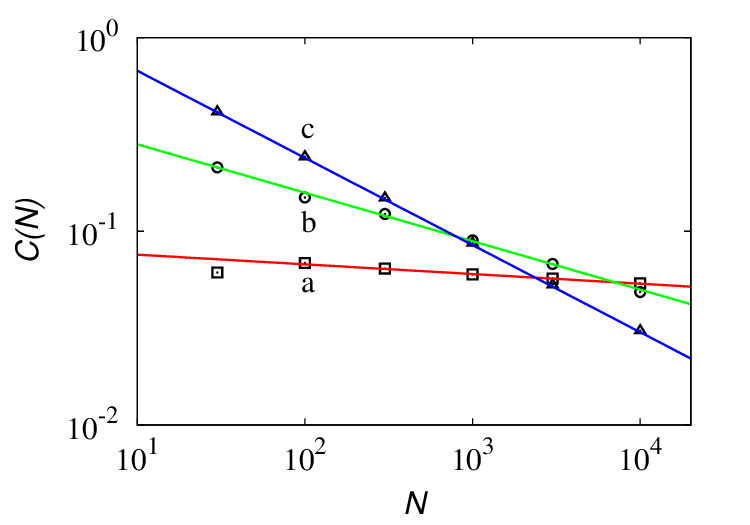}}
\end{center}
\caption{$C(N)$ vs.\ $N$ obtained for the random variable $z$
  defined by eq.~(\ref{eq5}) with $T(\tau)$ given by eq.~(\ref{eq3}),
  rescaled statistics are depicted in Fig.~\ref{Fig1}.  Symbols
  correspond to numerical results, lines represent
  eq.~(\ref{eq7}). Line (a) and ($\square$): $\xi=2.1$, line (b) and
  ($\circ$): $\xi=2.5$, line (c) and ($\triangle$): $\xi=2.9$.}
\label{Fig2}
\end{figure}

For the class of density
functions defined by eq.~(\ref{eq3}) one obtains
\begin{equation}
C(N) \sim N^{-(\xi-2)/2}
\label{eq7}
\end{equation}
as validated numerically in Fig.~\ref{Fig2}.
Observe that  eq.  (\ref{eq7}) implies
$C(N) \simeq C_0(\xi) N^{-(\xi-2)/2}$, where the prefactor $C_0(\xi)$ depends on
$\xi$ and is an increasing function of the exponent $\xi$. This
property can be justified in a rather intuitive way, by considering
the limit case $N=1$ at which $p(z;1)= [T(z)+T(-z)]/2$ for $z \in {\mathbb R}$,
where $T(z)=0$ for $z<0$, providing $p(z;1)= T(z)/2 \sim \xi/ 2 \, z^{\xi+1}$,
and thus a prefactor $C_0(\xi)$  increasing with $\xi$.
 However,
for $N$ sufficiently large, $C(N)|_{\xi_1}>C(N)|_{\xi_2}$, for $\xi_1<\xi_2$,
where $C(N)|_\xi$ is the value of the prefactor for the probability  density $T(\tau)$
eq. (\ref{eq3}) 
characterized by the value $\xi$.

The above results, namely eqs.~(\ref{eq6})-(\ref{eq7}), are consistent
with the uniform and local convergence theorems that have been
developed starting from the works by Berry \cite{berry} and Esseen
\cite{esseen}, known as the {\it Berry-Esseen theorem}.
Generalizations of the Berry-Esseen theorem have been thoroughly
addressed in the monograph by V.V.~Petrov \cite{petrov} (see
specifically Theorems 6 and 15 in \cite{petrov}),
and new estimates can be found in recent mathematical literature on
the subject \cite{nefedova}.  For the sake of precision,
the existing theorems reported in \cite{petrov} on local convergence
apply to the case $\xi>3$.
This is the reason why in
Fig.~\ref{Fig1} two cases, above and below $\xi=3$, have been
reported. The above analysis indicates that it would be possible to
extend these theorems also to the interval $\xi \in (2,3)$. 
This conjecture can be justified via a phenomenological argument (thus not a mathematical
proof)  developed below. 

Equations~(\ref{eq6}) and (\ref{eq7}) can be derived by applying the
classical characteristic function approach to the CLT.  Let
$\theta_h=\tau_h-\langle \tau_h \rangle$ and $\phi(k)=
\int_{-\infty}^\infty T(\theta) e^{i k \theta} d \theta$ the
characteristic function of the density $T(\theta)$. Due to the
power-law tail, it follows that
\begin{equation}
\phi(k) = 1 - \frac{k^2 \, {\sigma_\tau}^2}{2}
+ a \, |k|^\xi + O(k^{\xi+\varepsilon}) \quad ,
\label{eq_cf1}
\end{equation}
where $\sigma_\tau^2=\langle \tau^2 \rangle - \langle \tau \rangle^2$,
$a$ is a constant, and $O(k^{\xi+\varepsilon})$, $\varepsilon>0$,
 indicates terms of higher order with respect to $|k|^\xi$.
Since $z_N = \left ( \sum_{h=1}^N \theta_h \right )/\sqrt{N} \sigma_\tau$
and the variables $\theta_h$ are independent,
the characteristic function $\phi_{z_N}(k)=
\left [ \phi_\theta \left (\frac{k}{\sqrt{N} \sigma_\tau} \right ) \right ]^N$
is given by 
\begin{eqnarray}
\phi_{z_N}(k) & = & \left [ 1- \frac{k^2}{2 \, N} +   a  \left
( \frac{|k|}{\sqrt{N} \, \sigma_\tau} \right )^\xi \right ]^N \nonumber \\
&= & \left (1- \frac{k^2}{2 \, N} \right )^N + N \,
 \frac{a}{\sigma_\tau^\xi} \frac{|k|^\xi}{N^{\xi/2}} \, \left (1- \frac{k^2}{2 \, N} \right )^{N-1} + O(k^{2 \xi}) \quad 
\label{eq_cf2}
\end{eqnarray}
after omitting the higher-order terms.  In the high $z_N$-limit, i.e.,
$\xi>2$, $k^2 \ll 1$ and $N \gg 1$, from eq.~(\ref{eq_cf2}) it
follows that the leading terms are given by
\begin{equation}
\phi_{z_N}(k) \simeq
\exp{\left(-\frac{k^2}{2}\right)}\left(1+\frac{a |k|^\xi}{\sigma_\tau^\xi N^{\xi/2-1}}\right) \quad .
\label{eq_cf3}
\end{equation}
In the limit for $N$ tending to infinity eq.~(\ref{eq_cf3}) trivially
provides the normal distribution. Conversely, for large but finite
$N$ the second term within the parenthesis, corresponding to a
power-law scaling of the density $p(z;N)$ as $1/z^{\xi+1}$ with a
proportionality factor decaying as $N^{1-\xi/2}$, cf.\ to
eq.~(\ref{eq6}), is responsible for the divergence of the higher-order
moments. This result corresponds exactly to the singular limit
expressed by eq.~(\ref{eq5b}) for $N \rightarrow \infty$: The moments
of the limit distribution (Gaussian) are finite and do not coincide
with the limit for $N \rightarrow \infty$ of the moments of $p(z;N)$.

\subsection{Scaling analysis of the higher-order moments of Einsteinian L\'evy walks}
\label{sec2_01}

\begin{figure}[htbp]
\begin{center}
\resizebox{0.8\columnwidth}{!}{%
 \includegraphics{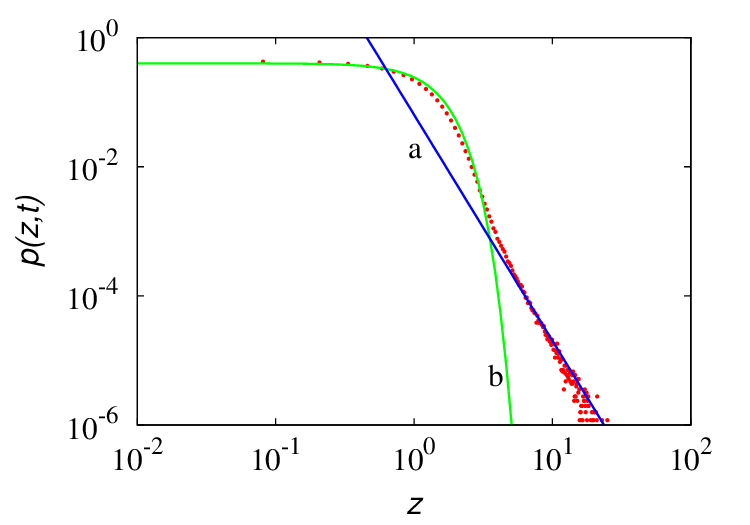}}
\end{center}
\caption{Probability density $p(z,t)$ vs.\ $z$ for
  $z=x(t)/\sigma_x(t)$ associated with a LW with $T(\tau)$ given by
  eq.~(\ref{eq3}) at $\xi=2.5$, $t=2000$.  Symbols refer to the result
  of stochastic simulations using $N_p= 2 \times 10^7$ realizations.
  Line (a) is proportional to $1/z^{1+\xi}$ while line (b) represents
  the normalized Gaussian distribution $g(z)$.}
\label{Fig2bis}
\end{figure}

Similar considerations apply for the LW eq.~(\ref{eq2}), 
because its kinematics is equivalent to the statistics of the sum of independent random variables 
(for $\xi>1$, such that $\langle \tau \rangle$ exists, and $t \simeq N \langle \tau \rangle$).  
This is demonstrated in Fig.~\ref{Fig2bis}, 
where we recall that $v_0=1$ a.u.\ and $x(0)=0$. 
In the case of the LW,
the random process $X(t)$ at time $t$ is bounded by the finite
propagation velocity $v_0$ so that $|x(t)|\leq t$.  Furthermore, $t
\sim N$, due to the boundedness of $\langle \tau \rangle$, implying $t
\sim N \, \langle \tau \rangle$, and eqs.~(\ref{eq6})-(\ref{eq7})
indicate that the tails of the probability density function $p(x,t)$
scale as
\begin{equation}
\sigma_x(t) \, p(x,t) \simeq \left . 
 \frac{\varepsilon^*(t)}{z^{\xi+1}} \right |_{z=x/\sigma_x(t)}=R(x,t) \, .
\qquad \varepsilon^*(t) \sim t^{-(\xi-2)/2}\quad ,
\label{eq8}
\end{equation}
This equation is defined for $|x| \in (a^*(t),t)$, where $a^*(t) <t$, 
and also satisfying the condition $\lim_{t \rightarrow \infty} a^*(t)=\infty$, 
is the crossover abscissa, associated with the transition from
the invariant Gaussian profile predicted by the CLT and the
large-deviation property defined by eq.~(\ref{eq8}).
Regarding the scaling of the even moments $\langle x^{2n}(t)\rangle$
(as for the odd moments $\langle x^{2 n +1}(t) \rangle =0$), the
approach used in \cite{lw_barkai} can be applied to eq.~(\ref{eq8})
leading to
\begin{equation}
\langle x^{2  n}(t) \rangle \sim t^{\gamma_n}
\, , \qquad \gamma_n = \max \{n, 2\, n+1-\xi \}
\label{eq9}
\end{equation}
In terms of the moment ratios (for $\xi >2$), one thus obtains
\begin{equation}
k_n(t) = \frac{\langle x^{2 n}(t) \rangle}{\langle x^2(t) \rangle ^n}
\sim t^{\alpha_n} \, ,  \qquad \alpha_n = \max \{0,n+1-\xi\}\quad .
\label{eq10}
\end{equation}
For $n=2$, $k_2(t)$ represents the kurtosis that, for an Einsteinian
LW with $\xi \in (2,3)$, deviates from the value predicted by the
limiting normal distribution, $k_2(t)=3$, as $k_2(t) \sim t^{3-\xi}$.
This effect is a consequence of the unbounded value of the third-order
moment of $T(\tau)$.

The same result can be predicted from the large-deviation properties
of the distribution $p(x,t)$ for high values of $t$ starting from
eq.~(\ref{eq8}).  Consider the effect of the long-tail of $R(x,t)$,
and set $ \langle x^{2 n}(t) \rangle_R= \sigma_x^{-1}(t) \int x^{2 n}
\, R(x,t) \, d x$. As above, consider an Einsteinian LW, i.e.,
$\xi>2$.  Since $R(x,t)$ is symmetric and different from zero in the
range $|x| \in (a^*(t),t)$, it follows from eq.~(\ref{eq8}) that
\begin{eqnarray}
\langle x^{2  n}(t) \rangle_R & \simeq &
 2 \,  
\int_{a^*(t)}^t  \frac{x^{2  n}}{\sigma_x(t)} \, R(x,t) \, d x  =
 \frac{2 \, \varepsilon^*(t)}{\sigma_x(t)}
 \,\int_{a^*(t)}^t x^{2  n} \, [x/\sigma_x(t)]^{-(\xi+1)} \, d x
\nonumber \\
& = & 2 \, \varepsilon^*(t) \, \sigma_x^\xi(t) \int_{a^*(t)}^t
x^{2 n - \xi -1} d x\quad .
\label{eq11}
\end{eqnarray}
Independently of the value of $a^*(t) < t$, the integral on the
r.h.s.\ of eq.~(\ref{eq11}) scales as $\int_{a^*(t)}^t x^{2 n - \xi
  -1} d x \sim t^{2n - \xi}$, and since $\sigma_x(t) \sim t^{1/2}$,
one finally obtains
\begin{equation}
\langle x^{2  n}(t) \rangle_R  \sim t^{-(\xi-2)/2} \, t^{\xi/2} \, t^{2 n -\xi}
\sim t^{2 n - \xi+1}\quad .
\label{eq12}
\end{equation}
Since $\langle x^{2 n}(t) \rangle= \langle x^{2 n}(t) \rangle_G+
\langle x^{2 n}(t) \rangle_R$, where $\langle x^{2 n}(t) \rangle_G$ is
the contribution due to the invariant Gaussian part of the probability
density,
\begin{equation}
\langle x^{2 n}(t) \rangle_G= \sigma_x^{-1}(t) \int_{-a^*(t)}^{a^*(t)}
x^{2 n} g(x/\sigma_x(t)) \, d x \sim t^n\quad ,
\label{eq11_bis}
\end{equation}
 eq.~(\ref{eq9}) follows, as the long-term moment scaling exponent,
 for fixed $n$, is the maximum value between $n$ and $2 \, n +1-\xi$.

\subsection{Examples elucidating the singular limit eq.~(\ref{eq5b}).}
\label{sec2bis}
As a first example, we consider 
the CTRW description of a LW parametrized with
respect to the operational time $N$, i.e., $x_N=\sum_{h=1}^N r_h \, \tau_h$,
in the case $\langle \tau^2 \rangle$ is finite.
Since $\langle x_N^2 \rangle   = N \, \langle \tau^2 \rangle$, and
\begin{equation}
\langle x_N^4 \rangle   =   N \, \langle \tau^4 \rangle + 3 \, \langle \tau^2
\rangle^2 \, N^2 ,
\label{eqad1}
\end{equation}
there are two cases to consider: 
(a) $\langle \tau^4 \rangle < \infty$. 
Thus, 
\begin {equation}
\lim_{N \rightarrow \infty} \frac{\langle x_N^4 \rangle}{\langle x^2_N \rangle^2} = 3 , 
\label{eqad2}
\end{equation}
corresponding to the kurtosis of the normal distribution. 
In this case, the equality is recovered in eq. (\ref{eq5b}) (at least for $n=2$).
(b) $\langle \tau^4 \rangle$ is unbounded. 
Consequently for any $N$
\begin{equation}
\langle x_N^4 \rangle = \infty \, , \qquad N=1,2,\dots
\label{eqad3}
\end{equation}
Eq. (\ref{eq5b}) applies strictly, as $3$ is
different from infinity. This clearly show in an unambiguous way the
singularity of the limit underlying the CLT in the presence of power-law tailed
distributions of the iid variables.
\footnote{It  is remarkable to observe the analogy between
CLT and renormalization group methods in field theory \cite{ren1,ren2}, in the
light of the singular limit eq. (\ref{eq5b}), corresponding to the
elimination of the unbounded singular terms, as those appearing in
eq. (\ref{eqad3}), by first performing  the ``renormalization'' with respect to
$N$,  corresponding to the r.h.s in eq. (\ref{eq5b}), eliminating in this way 
the singularities
that may occur for power-law tailed distributions. }

As a second example, 
we consider a LW in a continuous time setting. 
The main conceptual difference with respect to the CTRW description 
is that for any $t$ the moments $\langle x^{2n}(t) \rangle$
are bounded for any $t>0$ due to the finite propagation velocity. Since $v_0=1$ a.u., we have $\langle x^{2 n}(t) \rangle \leq t^{2 n}$.
As regards the relation between the physical  $t$ and the operational time $N$
we have
\begin{equation}
t = N \, \langle \tau \rangle \, \left ( 1+ O \left ( \frac{1}{\sqrt{n}} \right ) \right )
\label{eqad4}
\end{equation}
In the present analysis  the higher-order contribution $O(n^{-1/2})$ is completely
 irrelevant, and for the sake of notational simplicity it will be omitted.
We can still apply eq. (\ref{eqad1}) but we have to consider that at time $t$ the
density function for $\tau$ is not $T(\tau)$ but the censored distribution
$T_t(\tau)$, where
\begin{equation}
T_t(\tau) =
\left \{
\begin{array}{lll}
T(\tau)  & \;\; & \tau<t \\
\left (1- F(t)  \right )   \, \delta(\tau-t) & \;\; & \tau \geq t
\end{array}
\right .
\label{eqad5}
\end{equation}
where $F(t)= \int_0^t T(\tau) \, d \tau$.
Consequently, eq. (\ref{eqad1}),  for a LW in a continous time frame, reads 
\begin{equation}
\langle x^4(t) \rangle   =   \frac{t \, \langle \tau^4 \rangle_t}{\langle \tau \rangle_t}
 + \frac{3 \,  t^2 \, \langle \tau^2
\rangle_t^2}{\langle \tau \rangle_t^2} 
= \langle x^4_{\rm ecc}(t) \rangle + \frac{3 \,  t^2 \, \langle \tau^2
\rangle_t^2}{\langle \tau \rangle_t^2}
\label{eqad6}
\end{equation}
where for any function $f(\tau)$, $\langle f(\tau) \rangle_t= \int_0^\infty f(\tau) \, T_t(\tau)$. Thus,
\begin{equation}
\langle x^4_{\rm ecc}(t) \rangle =  
\frac{t}{\langle \tau \rangle} \int_0^\infty \tau^4 \, T_t(\tau) \, d \tau
= \frac{t}{\langle \tau \rangle} \, I(t,4,\xi) + \frac {t}{\langle \tau \rangle}
\frac {t^4}{(1+t)^\xi}
\label{eqad7}
\end{equation}
where
\begin{equation}
\hspace{-2cm} I(t,n,\xi)= \xi \int_0^t \frac{\tau^n \, d \tau}{(1+\tau)^{\xi+1}}
=
\left \{
\begin{array}{lll}
- \frac{t^n}{(1+t)^\xi} + \frac{n}{\xi-1} \, I(t,n-1,\xi-1)  & \;\;  & n \geq 1 
\\
1- \frac{1}{(1+t)^\xi} & \;\; & n=0
\end{array}
\right .
\label{eqad8}
\end{equation}
It follows from eqs. (\ref{eqad7})-(\ref{eqad8}) that
\begin{equation}
\langle x^4_{\rm ecc}(t) \rangle \sim t^{5-\xi}
\label{eqad9}
\end{equation}
and more generally, using the same approach,
\begin{equation}
\langle x^{2 n}(t) \rangle  \sim a \, t^{2 n} + b  \,t^{2  n +1 -\xi}
\label{eqad10}
\end{equation}
where $a,b>0$ are constant, consistently with the  scaling theory of the
probability density functions at finite $N$ developed previously.

A third analytical example, exploring the singular
  limit for non identically distributed random variables, is
  included in Appendix~A.

\section{Moments of LWs: hyperbolic statistical description}
\label{sec3new}

A convenient way for approaching the statistical characterization of
LWs is to use the hyperbolic partial density formalism
envisaged by Fedotov \cite{fedotov}, applied in
\cite{fedotov1,fedotov2} and further elaborated by Giona et
al.\ \cite{klages}. For a LW, the partial probability densities
describing the process, $p_\alpha(x,t;\tau)$, are parametrized with
respect to the velocity direction ($\alpha=1,2$) and the
transitional age $\tau \in [0,\infty)$, representing the time elapsed
  from the latest velocity transition. The partial densities satisfy
  the equations
\begin{equation}
\frac{\partial p_\alpha(x,t;\tau)}{\partial t}= - b_\alpha \, \frac{\partial p_\alpha(x,t;\tau)}{\partial x} - \frac{\partial p_\alpha(x,t;\tau)}{\partial \tau} - \lambda(\tau) \,
p_\alpha(x,t;\tau)\quad .
\label{eq_x1}
\end{equation}
Here $\lambda(\tau)$ is the transition rate, which is related to the
transition time density $T(\tau)$, characterizing the CTRW formulation,
by the equation
\begin{equation}
T(\tau)= \lambda(\tau) \, \exp \left [-\int_0^\tau \lambda(\tau^\prime)
\, d \tau^\prime \right ] \quad .
\label{eq_x1bis}
\end{equation}
 In the case of eq.~(\ref{eq3}), $\lambda(\tau)=\xi/(\tau_0+\tau)$.
 In eq.~(\ref{eq_x1}) the velocities of the two partial waves are
 moving in opposite directions, i.e., $b_1=v_0=1$, $b_2=-v_0=-1$.
 Equations~(\ref{eq_x1}) are equipped with the boundary condition
 accounting for the velocity transitions
\begin{equation}
p_\alpha(x,t;0)= \frac{1}{2} \sum_{\beta=1}^2 \int_0^\infty 
\lambda(\tau) \, p_\beta(x,t;\tau) \ d \tau\quad .
\label{eq_x2}
\end{equation}
The overall probability density function for the position of a L\'evy
walker at time $t$ is thus $p(x,t)=\sum_{\alpha=1}^2 \int_0^\infty
p_\alpha(x,t;\tau) \, d \tau$.  From eqs.~(\ref{eq_x1}) and (\ref{eq_x2}),
the equations for the partial moment hierarchy
$m_\alpha^{(n)}(t,\tau)= \int_{-\infty}^\infty x^n \,
p_\alpha(x,t;\tau) \, d x$, $n=0,1,2,\dots$, follow as
\begin{equation}
\frac{\partial m_\alpha^{(n)}(t,\tau)}{\partial t}
= - \frac{\partial m_\alpha^{(n)}(t,\tau)}{\partial \tau} + n \, b_\alpha
\, m_\alpha^{(n-1)}(t,\tau) - \lambda(\tau) \, m_\alpha^{(n)}(t,\tau)
\label{eq_x3}
\end{equation}
with
\begin{equation}
m_\alpha^{(n)}(t,0) = \frac{1}{2} \sum_{\beta=1}^2 \int_0^\infty
\lambda(\tau) \, m_\beta^{(n)}(t,\tau) \, d \tau\quad ,
\label{eq_x4}
\end{equation}
and the overall $n$-th order moment is expressed by
\begin{equation}
M^{(n)}(t)= \sum_{\alpha=1}^2 \int_0^\infty m_\alpha^{(n)}(t,\tau) \, d \tau\quad .
\label{eq_x41}
\end{equation}
If initially all the particles are located at the origin, $x=0$,
possess a transitional age $\tau=0$ (corresponding to a full
``transitional synchronization'' of the walkers' ensemble), and their
velocities are distributed amongst the two velocity directions in an
equiprobable way, the initial conditions for the moment dynamics
eqs.~(\ref{eq_x3}) and (\ref{eq_x4}) read $m_\alpha^{(0)}(0,\tau)=
\delta(\tau)/2$ and $m_\alpha^{(n)}(0,\tau)=0$ for $n=1,2,\dots$.

\begin{figure}[htbp]
\begin{center}
\resizebox{0.8\columnwidth}{!}{%
 \includegraphics{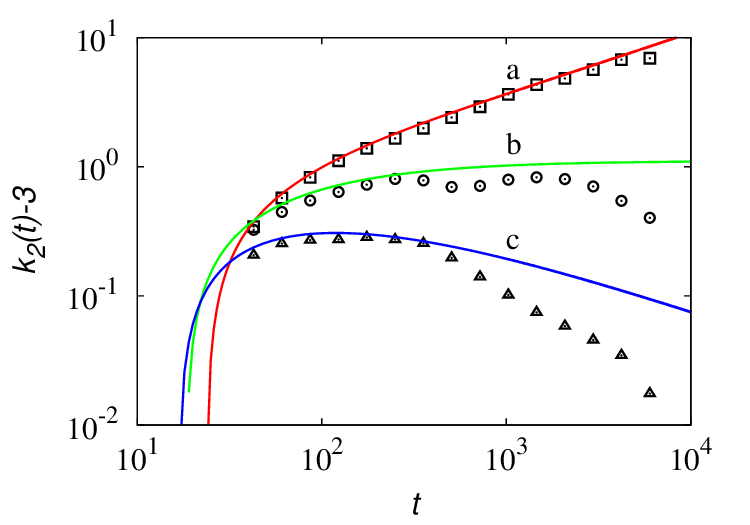}}
\end{center}
\caption{Deviation of the kurtosis $k_2(t)$ from the normal value
  $k_{2,{\rm norm}}=3$ vs $t$, for LWs defined by $T(\tau)$ given by
  eq.~(\ref{eq3}), at several values of $\xi$. Symbols represent the
  results of stochastic simulations using an ensemble of $N_p=10^7$
  realizations, lines are the solutions of the moment equations
  (\ref{eq_x3})-(\ref{eq_x4}). Line (a) and ($\square$): $\xi=2.5$,
  line (b) and ($\circ$): $\xi=3$, line (c) and ($\triangle$):
  $\xi=3.5$. }
 \label{Fig3}
 \end{figure}

The moment equations (\ref{eq_x3}) and (\ref{eq_x4}) can be solved
numerically, and their scaling in time can also be derived
analytically \cite{klages}, at least for the lower-order elements of
the moment hierarchy. Regarding the numerics, since the ``velocity of
propagation through the age'' is constant, $d\tau/dt=1$, a finite
difference approach can be employed, setting the time step $\Delta t$
equal to the age discretization $\Delta \tau$ in order to avoid spurious
effects associated with numerical diffusion.  Figure \ref{Fig3}
depicts the behaviour of the difference between the kurtosis $k_2(t)$
and its normal value, $k_{2,{\rm norm}}=3$, as a function of time $t$ 
for three different values of $\xi$ (solid lines): $\xi=2.5$, leading to an
anomalous fourth-order behavior; $\xi=3$, corresponding to the
threshold value between anomalous (below) and normal (above) kurtosis;
and $\xi=3.5$ above the threshold. 
These results are obtained by solving numerically the moment equations (\ref{eq_x3}) and (\ref{eq_x4}),

As expected, $k_2(t)$ diverges to infinity at $\xi=2.5$, while for
$\xi=3.5$, $k_2(t)-3$ relaxes asymptotically to zero.  The threshold
value $\xi=3$ is also interesting, as from eq.~(\ref{eq10}), $k_2(t)$
should at most increase with time slower than any power $t^\eta$,
$\eta>0$.  Numerical data indicates that at the threshold $\xi=3$,
$k_2(t)$ attains, for long times, values manifestly different form $k_{2,{\rm norm}}$. 
In this case, whether the kurtosis would eventually converge towards a constant limit value 
or increase in a non-power law (possibly logarithmic) way cannot be determined from this analysis.

These predictions are further validated by numerical simulation of the stochastic dynamic (markers).  The data are obtained by simulating an ensemble
of $N_p=10^7$ L\'evy walkers initially located at $x=0$, with
$\tau=0$ and equiprobable velocity directions. The stochastic
simulations qualitatively agree with the solution of the moment
equations, and a consistent quantitative agreement is achieved for $t
\leq 5 \times10^2$. As the time increases, the statistical errors
associated with the finite size of the ensemble become significant. In
order to reach an accurate prediction of the fourth-order moments, the
size of the ensemble should be increased beyond the limit considered
here, $N_p=10^7$, by several orders of magnitudes.  This comparison,
incidentally, shows the advantages on the use of statistical moment
equations (\ref{eq_x3}) for addressing the large-deviation properties
of LWs with respect to the direct stochastic simulation of the
dynamics.

\begin{figure}[htbp]
\begin{center}
\resizebox{0.8\columnwidth}{!}{%
 \includegraphics{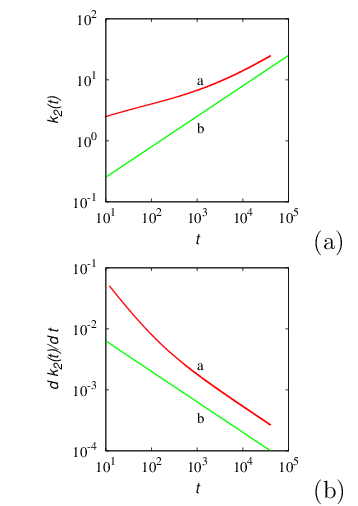}}
\end{center}
\caption{Kurtosis $k_2(t)$ vs.\ $t$ for a LW at $\xi=2.5$ (panel a)
  and its time derivative (panel (b)). Lines (a) in panels (a) and (b)
  represent $k_2(t)$ and $d k_2(t)/d t$, respectively, obtained from
  the numerical solution of the moment equations
  (\ref{eq_x3})-(\ref{eq_x4}), lines (b) in panels (a) and (b) depict
  the scalings $t^{1/2}$ and $t^{-1/2}$, respectively.}
 \label{Fig4}
 \end{figure}

 \ref{Fig4} depicts a simulation for long times of the
scaling of the kurtosis $k_2(t)$ at $\xi=2.5$, from which
it is possible to estimate the scaling exponent. Numerical
simulation based on the moment equations provide a confirmation
of the theoretical scaling $k_2(t) \sim t^{1/2}$ predicted by
eq.~(\ref{eq10}) at $\xi=2.5$.

\section{Thermodynamic implications}
\label{sec3}
Equation~(\ref{eq9})
implies interesting thermodynamic
consequences for its statistical interpretion. 
Focusing primarily on the superdiffusive case $\xi \in (1,2)$, 
several authors have pointed out that 
the scaling relation~(\ref{eq9}) corresponds to the manifestation of strong
anomalous properties \cite{voll,lw_barkai},
because the family of even order moments $M^{(2
  n)}(t)$ does not scale with a unique exponent $\gamma_1$, i.e.,
$M^{(2 n)}(t) \neq t^{n \gamma_1}$. 
This behavior indeed cannot find a
statistical interpretation in terms of a single scaling function
$F(z)$ for the overall density function $p(x,t)$, i.e., 
this distribution cannot be expressed in the form of
$p(x,t)= t^{-\gamma_1/2} F(x/t^{\gamma_1/2})$ \cite{lw3}.  Viewed in the broader
perspective of generic values $\xi>1$, thus including 
regular diffusive LWs, the statistical implications of Eq.~(\ref{eq9}) 
can be interpreted differently. 
While it is true that for the superdiffusive case $\xi\in (1,2)$ 
$\gamma_n \neq n \, \gamma_1$, Eq.~(\ref{eq9}) indicates that 
there exists just a single exponent branch $\gamma_n= 2 n + 1 -
\xi$ for any $n=1,2,\dots$, at least for the even-order classical
moments $M^{(2 n)}(t)$.  This observation will be further adressed
below.  Conversely, in the Einstenian case $\xi>2$, the
scaling of the whole moment hierarchy derives from the combination of
two different effects, thus producing the crossover behavior
\begin{equation}
\gamma_n= \left \{
\begin{array}{lll}
n & \;\; \; & \mbox{for} \; n < \xi-1 \\
2 \, n +1 -\xi & \;\;\; & \mbox{for} \; n> \xi-1
\end{array}
\right .
\label{eq5_1}
\end{equation}
for fixed $\xi$. The lower-order elements of the moment hierarchy, up
to $n<\xi-1$ are controlled by a regular diffusive scaling, while
higher-order moments are subject to the same scaling that
homogeneously applies to the superdiffusive case for even-order
moments. In this regard, the physical phenomenology occurring for
$\xi>2$, viewed in the light of the whole moment hierarchy, appears to
be ``more heterogeneous'' than that observed for $\xi \in (1,2)$.

The physical reason for this crossover behavior can be traced back to 
the propagation mechanism characterizing LWs,
made evident by the hyperbolic formulation.  
The overall dynamics of a LW is controlled by: (i) the
recombination of the partial density waves associated with the two
different velocity directions at finite values of the transition age
$\tau$, and (ii) the existence of a statistically significant fraction
of particles never performing recombination and thus propagating
ballistically.  The first mechanism determines the regular diffusive
scaling of eq.~(\ref{eq5_1}) for $n<\xi-1$; 
the second one
controls the other scaling that
emerges for higher values of $n$. The latter observation becomes
evident by observing that, for $n>\xi-1$, the difference
$\gamma_{n+1}-\gamma_n=2$ is purely ballistic. The interplay between
the two mechanisms is present for any value of $\xi$. While for $\xi
\in (1,2)$ the ballistic propagation overwhelms the regular diffusive
scaling for any even-order moments, values of $\xi>2$ permit to
appreciate the simultaneous presence of the two mechanisms.

From this physical interpretation, it follows that it is almost
impossible to derive a single evolution equation for the overall
probability density function $p(x,t)$, which is local in time by not
requiring memory integrals involving the whole previous statistical
history \cite{fedotov}, providing the correct scaling of the whole
moment hierarchy, because it would involve the interplay between two
completely different propagation mechanisms. Nonetheless, the analysis
also suggests a different, physically simple approach towards the
thermodynamics of LW processes.  It consists of considering a LW
ensemble as being made of two phases: a diffusionally recombining
phase, and a purely ballistic one. This approach is altogether similar
to the classical Landau treatment of superfluidity
\cite{superfluidity}, where a viscous and a purely inviscid phase
coexist. It should be mentioned that models for LWs involving
fractional derivative operators have been proposed in the literature
\cite{taylor-king}, describing higher-order anomalies with
time-dependent exponents \cite{dossantos}. The resulting macroscopic
properties, expressed by the scaling of the whole moment hierarchy,
derive from the recombination of the two phases. If the anomalous
scaling occurring for $n>\xi-1$ is the macroscopic consequence of the
presence of two non-recombining ballistic waves moving in the two
opposite directions, the ballistic phase can be simply described by
means of two impulsive waves $P_\pm(x,t)$ moving at constant velocity
$v_0$,
\begin{equation}
P_\pm(x,t)= \frac{\Phi(t)}{2} \, \delta(x \mp v_0\,t)\quad .
\label{eq5_2}
\end{equation}
The key quantity here is the survival function $\Phi(t)$ \cite{lw3}.  At time
$t=0$, $\Phi(0)=1$, and as time $t$ proceeds, particles leave the
ballistic phase to join the diffusive one.  The function $\Phi(t)$
does not coincide, in this approximate statistical description, to the
bare survival probability $\Phi_0(t)=e^{-\Lambda(t)}$, where
$\Lambda(t)=\int_0^t \lambda(\tau) \, d \tau$ (for
$\lambda(\tau)=\xi/(1+\tau)$, $\Phi_0(t)=(1+t)^{-\xi}$), for the
reason that a continuous flow of particles from the diffusive phase
contributes to the ballistic one, and this is the reason why the
ballistic moment exponent equals $\gamma_n=2 \, n +1 -\xi$, and not $2
\, n -\xi$.  This contribution can be modelled by considering
$\Phi(t)$ to be proportional to the product of $\Phi_0(t)$ and of the
elapsed time $t$, i.e., $\Phi(t) \sim t \, \Phi_0(t)$. For this
reason, and in order to fulfill the initial condition, we set
$\Phi(t)=(1+t) \, \Phi_0(t)=(1+t)^{-(\xi-1)}$, so that $\Phi(0)=1$,
and $\lim_{t\rightarrow \infty} \Phi(t)=0$.

\begin{figure}[htbp]
\begin{center}
\resizebox{1.0\columnwidth}{!}{%
 \includegraphics{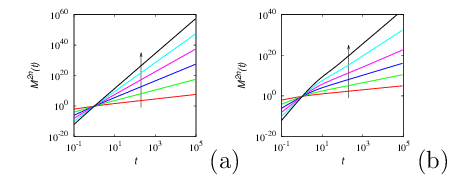}}
\end{center}
\caption{Scaling of the global moment hierarchy $M^{2n}(t)$ vs.\ $t$
  obtained from the approximated ``two-phase'' hyperbolic model
  eqs.~(\ref{eq5_5}) and (\ref{eq5_6}).  Panel (a) refers to
  $\xi=1.5$, panel (b) to $\xi=4.5$.}
\label{Fig5}
\end{figure}

\begin{figure}[htbp]
\begin{center}
\resizebox{0.8\columnwidth}{!}{%
 \includegraphics{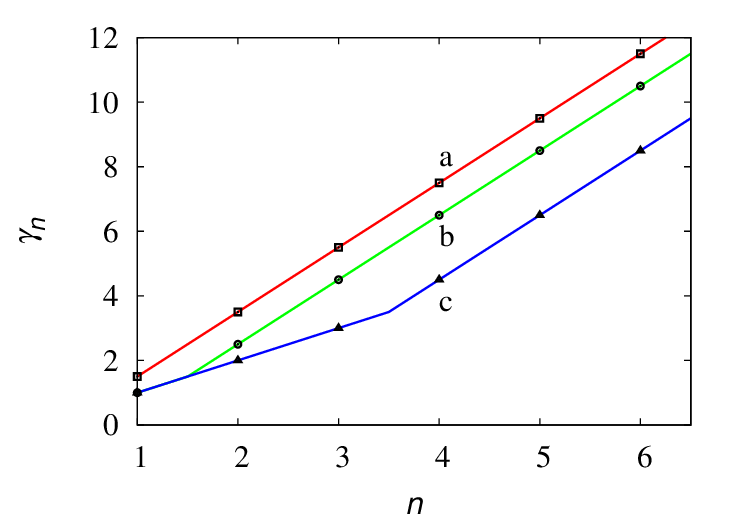}}
\end{center}
\caption{Moment exponents $\gamma_n$ vs.\ $n$. Lines are the
  theoretical predictions eq.~(\ref{eq9}), symbols the results derived
  from the approximate ``two-phase'' hyperbolic model
  eqs.~(\ref{eq5_5}) and (\ref{eq5_6}).  Line (a) and ($\square$)
  refer to $\xi=1.5$, line (b) and ($\circ$) to $\xi=2.5$, line (c)
  and ($\triangle$) to $\xi=4.5$.}
\label{Fig6}
\end{figure}

Next, we consider the diffusing phase, and let $q_\pm(x,t)$ be the
associated partial density waves. Assuming in this approximate
statistical model a constant non zero-value $\lambda_q>0$ for the
transition rate amongst forward/backward propagating partial waves
of the diffusing phase, the evolution equations for $q_\pm(x,t)$
coincide with those associated to a Poisson-Kac process
\cite{kac,gpk1}, onto which a continuous flux of particles from the
ballistic phase is superimposed. The latter contribution can be
derived from probability conservation and, assuming symmetric
conditions in the recombination amongst forward and backward
propagating waves, it equals $k(t) \, \left [ \delta(x-v_0 t) +
  \delta(x_0+v_0 \, t) \right ]/4$, where
\begin{equation}
k(t)= - \frac{d \Phi(t)}{d t}  \geq 0\quad .
\label{eq5_3}
\end{equation}
Consequently, the balance equation for $q_\pm(x,t)$ becomes
\begin{eqnarray}
\frac{\partial q_\pm(x,t)}{\partial t} & = & \mp v_0 \, \frac{\partial q_\pm(x,t)}{\partial x}  \mp \lambda_q \, \left  [q_+(x,t)-q_-(x,t) \right ]
\nonumber \\
&+& \frac{k(t)}{4} \,  \left [ \delta(x-v_0 t) + \delta(x_0+v_0 \, t) \right ]
\label{eq5_4}
\end{eqnarray} 
with the  initial condition $q_\pm(x,0)=0$.
Observe from eq.~(\ref{eq5_2}) the symmetry in the dynamics of the
ballistic and diffusing phase, as $P_\pm(x,t)$ satisfy 
the system of equations
\begin{equation}
\frac{\partial P_\pm(x,t)}{\partial t}= \mp  v_0 \,\frac{\partial P_\pm(x,t)}{\partial x}
- \lambda_b(t) \, P_\pm(x,t)\quad ,
\label{eq5_5}
\end{equation}
where
\begin{equation}
\lambda_b(t)= - \frac{1}{\Phi(t)} \frac{d \Phi(t)}{d t}\quad ,
\label{eq5_5bis}
\end{equation} 
for instance $\lambda_b(t)=(\xi-1)/(1+t)$ for
$\Phi(t)=(1+t)^{-(\xi-1)}$, equipped with the initial conditions
$P_\pm(x,0)=\delta(x)/2$. Then eq.~(\ref{eq5_4}) can be rewritten as
\begin{equation}
\hspace{-1.6cm} \frac{\partial q_\pm(x,t)}{\partial t} = \mp v_0 \, \frac{\partial q_\pm(x,t)}{\partial x}  \mp \lambda_q \, \left [ q_+(x,t)-q_-(x,t) \right ]
+ \frac{\lambda_b(t)  }{2} \left [ P_+(x,t) +P_-(x,t) \right ]\, .
\label{eq5_6}
\end{equation}
Both these phases propagate ballistically, with the major difference
that a continuous recombination between $q_\pm(x,t)$ occurs between
the two partial waves of the diffusive phase, determining the long-term diffusive
behavior, which is absent in the dynamics of the ballistic phase.
Therefore the ballistic phase is a deterministically propagating
phase without internal recombination dynamics, subject solely
to a flux from and to the diffusive phase.
In this sense, the analogy with the multiphase description of superfluidity
is strict \cite{superfluidity}.
The system of equations  (\ref{eq5_5}) and (\ref{eq5_6}) conserves probability,
\begin{equation}
\sum_{\alpha=\pm} \int_{-\infty}^\infty \left [ q_\alpha(x,t)+
P_\alpha(x,t) \right ]  d x = 1\quad ,
\label{eq5_7}
\end{equation}
and $q_\alpha(x,t)\geq 0$, $P_\alpha(x,t) \geq 0$.  Let
$m_{\pm}^{(n)}(t)=\int_{-\infty}^\infty x^n \, q_\pm(x,t) \, d x$ be
the moments associated with the diffusing phase, satisfying the
balance equations
\begin{equation}
\hspace{-1.6cm} 
\frac{d m_\pm^{(n)}(t)}{d t} = \pm n \, v_0 \, m_\pm^{(n-1)}(t)
\mp \lambda_q \, \left [ m_+^{(n)}(t)-m_-^{(n)}(t) \right ]
+ \frac{k(t) \, (v_0 \, t)^n}{4} \, \left [1-(-1)^n   \right ]\,.
\label{eq5_8}
\end{equation}
The overall moments $M^{(n)}(t)$ of the process are the sum
of the partial moments of the diffusing and ballistic phases,
\begin{equation}
M^{(n)}(t)= m_+^{(n)}(t)+ m_-^{(n)}(t) + \frac{\Phi(t) \, (v_o \, t)^n}{2} \, \left [1+(-1)^n
\right ]\quad .
\label{eq5_9}
\end{equation}
It follows from the symmetries that the odd moments $M^{(2 n+1)}(t)=0$
are identically vanishing. Figure~\ref{Fig5} depicts the scaling of
the even moments $M^{(2 n)}(t)$ for $n=1,\dots,6$ 
for $\xi=1.5$ (panel a) and $\xi=4.5$ (panel b), obtained from
the numerical integration of the moment equations~(\ref{eq5_8}). 
Once again, we have set $v_0=1$ a.u., and $\lambda_q=1$ a.u.
The values of the long-term moment scaling exponents $\gamma_n$
derived from this data are depicted in Fig.~\ref{Fig6}. They perfectly
agree with the theoretical result expressed by eq.~(\ref{eq9}) for
$\xi>1$.

The two-phase model eqs.~(\ref{eq5_5}) and (\ref{eq5_6}) has been
essentially derived for higher-order anomalous (superdiffusive) LWs,
and the agreement between its predictions and the correct scaling of
the even-order moments $M^{(2n)}(t)$ in the anomalous case, i.e., for
$1<\xi<2$, is just a fortuituous case, as addressed in the next
paragraph.  The result for $\gamma_n$ can now be also recovered from
the long time limit of eq.~(\ref{eq5_4}), which in the present case
corresponds to the Kac limit (parabolic approximation) of this
Poisson-Kac process: Indicating with $q(x,t)=q_+(x,t)+q_-(x,t)$ the
overall density function of the diffusing phase, $q(x,t)$ satisfies in
the long time limit the diffusion equation
\begin{equation}
\frac{\partial q(x,t)}{\partial t}= D_q \, \frac{\partial^2 q(x,t)}{\partial x^2}+ \frac{k(t)}{4} \, 
\left [ \delta(x-v_0 t) + \delta(x+v_0 \, t) \right ]\quad ,
\label{eq5_10}
\end{equation}
where $D_q=v_0^2/2 \,\lambda_q$, from which a closed-form expression
for the moment scaling exponents $\gamma_n$ follows 
that enforces the definition~(\ref{eq5_3}) of $k(t)$.

\subsection{Generalized moments, differential moment exponents
 and anomalous behavior}
\label{sec3_1}

Albeit the two phase model described in the previous paragraph
correctly reproduces the scaling of the moment hierarchy
$\{M^{(n)}(t)\}_{n=0}^\infty$ for $\xi>1$, its physical application
shuold be properly limited to the case $\xi>2$, i.e., to the case of
higher-order anomalous diffusion. The reason is that it predicts a
Gaussian shape for the overall density function near $x=0$, and this
is correct solely for $\xi>2$. Another way to look at this problem is
to consider the generalized moments $\langle |x(t)|^q\rangle$, with $q
\in {\mathbb R}^+$, that provide not only a way to detect the
occurrence of strong anomalous properties, but also to probe the
physical mechanism of the diffusive propagation.

\begin{figure}[htbp]
\begin{center}
\resizebox{0.8\columnwidth}{!}{%
 \includegraphics{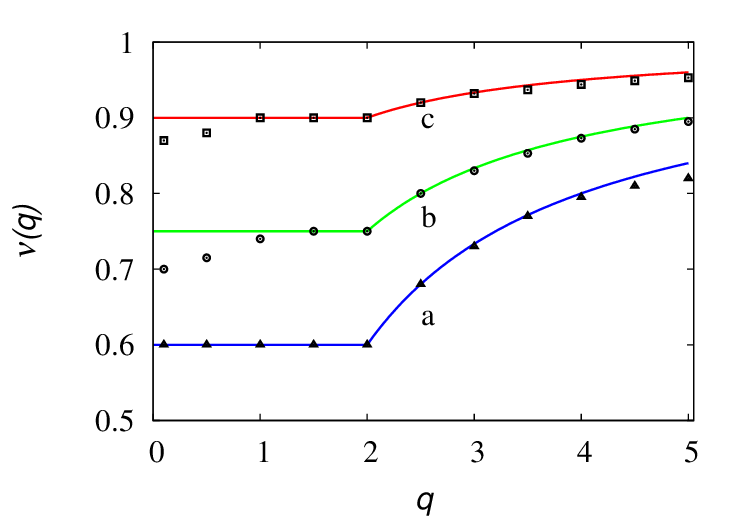}}
\end{center}
\caption{Moment scaling exponent $\nu(q)$ vs.\ $q$ for anomalous
  superdiffusive LWs characterized by the transition time density
  eq.~(\ref{eq3}) with $\tau_0=1$, $v_0=1$. Symbols are the results of
  stochastic simulations, lines represent eq.~(\ref{eq5_13}). Line (a)
  and ($\triangle$) refer to $\xi=1.8$, line (b) and ($\circ$) to
  $\xi=1.5$, line (c) and ($\square$) to $\xi=1.2$.}
\label{Fig7}
\end{figure}

In the higher-order anomalous regime $\xi>2$, the exponent spectrum
$\nu(q)$ defined by eq.~(\ref{eq1_1}) is a non-trivial function of
$q$,
\begin{equation}
\nu(q)= \left \{
\begin{array}{lll}
1/2 & \;\; & \mbox{for} \;  q< q_c  \\
1-(\xi-1)/q & \;\; & \mbox{for} \; q> q_c 
\end{array}
\right .\quad ,
\label{eq5_16}
\end{equation}
where the crossover order $q_c=2 \, (\xi-1)$ depends on $\xi$.  This
crossover is the signature of the change of regime between the two
propagation mechanisms described in the previous paragraph.  Let
$\gamma^*(q)=q \, \nu(q)$, so that $\langle |x(t)|^q \rangle \sim
t^{\gamma^*(q)}$.  The phenomenon of higher-order anomalous behavior
can be described in an even more convenient way by introducing the
differential moment exponents $\nu^*(q)$ defined as
\begin{equation}
\nu^*(q)= \frac{d \gamma^*(q)}{d q}\quad .
\label{eq5_11}
\end{equation}
The spectrum of differential exponents $\nu^*(q)$ is related to
$\nu(q)$ by the equation $\nu^*(q)= \nu(q)- q \, d \nu(q)/d q$.  
It follows therefore 
that $\nu^*(q)$ attains two constant values independent on $q$ in the two regimes, i.e., 
\begin{equation}
\nu^*(q)= \left \{
\begin{array}{lll}
1/2 & \;\; & \mbox{for} \;  q< q_c \\
1 & \;\; & \mbox{for} \; q> q_c
\end{array}
\right . \quad ,
\label{eq5_12}
\end{equation}
thus corresponding to the diffusive and ballistic propagation mechanisms.
Let us now consider the anomalous superdiffusive case, $\xi \in (1,2)$. The
behavior of the moment exponents $\nu(q)$ is given by
\begin{equation}
\nu(q)= \left \{
\begin{array}{lll}
(3-\xi)/2 & \;\; & \mbox{for} \;  q< 2  \\
1-(\xi-1)/q & \;\; & \mbox{for} \; q> 2 
\end{array}
\right .\quad ,
\label{eq5_13}
\end{equation}
as depicted in Fig.~\ref{Fig7}. Shown therein are results from
stochastic simulations performed by considering an ensemble of $10^6$
particles and evaluating the long-term scaling of the generalized
moments.
The slight disagreement between theory and simulations for low values
of $q<1$ is due to numerical errors, in view of the relatively small
size of the ensemble considered.  The remarkable property associated
with eq.~(\ref{eq5_13}) is that the crossover order $q_c$ does not
depend on $\xi$ and is equal to $2$. This is the reason why it is not
possible to detect any transition between the two propagation
mechanisms from the hierarchy of integer moments $M^{(n)}(t)$,
$n=0,1,2$, as $M^{(0)}(t)=1$ by probability conservation, and the
first-order moment is at most constant, depending on the initial
preparation of the particle system.

In terms of the differential exponents $\nu^*(q)$, eq.~(\ref{eq5_13}) yields
\begin{equation}
\nu^*(q)= \left \{
\begin{array}{lll}
(3-\xi)/2 & \;\; & \mbox{for} \;  q< 2 \\
1 & \;\; & \mbox{for} \; q> 2
\end{array}
\right . \quad .
\label{eq5_14}
\end{equation}
This corresponds to an
effective anomalous diffusion regime characterized by the scaling law
$\langle | x(t) | \rangle \sim t^{(3-\xi)/2}$, representing the
lower-order part of the generalized moment hierarchy, and to a
ballistic motion that becomes evident for $q>2$.  The properties of
the lower-order part of the generalized moment hierarchy clearly show
that the approximate model developed in the previous paragraph cannot
be extended beyond the range of higher-order anomalous behavior, as it
predicts for the generalized moments at low $q$ values a regular
diffusive scaling for any $\xi$, instead of the correct one $\langle
|x(t)|^q \rangle \sim t^{q \, (3-\xi)/2}$.

\section{Conclusions}
\label{sec5}

In this article we have analyzed the statistical properties of regular
Einsteinian LWs characterized by a power-law tail $T(\tau) \sim
\tau^{-(1+\xi)}$, $\xi>2$, in the transition-time density function,
using the CLT as a microscope to detect the peculiarities in this
class of stochastic propagation processes.  While the mean square
displacement in the long-term limit scales linearly as a function of
time, for any $\xi>2$, there exists an integer $n^*$ such that $M^{(2
  n)}(t)/t^n \rightarrow \infty$ for $n>n^*$ and $t\rightarrow
\infty$. This observation has led to the concept of higher-order
anomalous diffusion, and to a finer characterization of its
propagation properties by considering the hierarchy of generalized
moments $\langle |x(t)|^q \rangle$, $q \in {\mathbb R}^+$.  In this
perspective, any LW characterized by a power-law tail in $T(\tau)$ is
intrinsically anomalous with respect to the overall statistical
properties expressed by the whole moment hierarchy.  For transitionally
ergodic LWs ($\xi>1$) this phenomenon is essentially a consequence of
two distinct transport mechanisms: (i) a diffusive scaling (be it
regular for $\xi>2$, or anomalous for $1 < \xi < 2$) described by the
scaling of the lower-order elements of the generalized moment
hierarchy, and by (ii) a ballistic propagation deriving from the
statistical occurrence of a significant fraction of particles never
performing transitions in the velocity direction, observable for
higher values of $q$.  Viewed from the perspective of the properties
of the moments, Einsteinian LWs possessing power-law tails in the
density $T(\tau)$ as in eq.~(\ref{eq3}) provide a higher level of
heterogeneity than anomalous LWs with $\xi<2$.  For instance, as
remarked in \cite{rebenshtok16}, the knowledge of the infinite density
\cite{lw4,lw_barkai} is not sufficient to achieve a complete prediction of the
moment hierarchy, and this phenomenon has been attributed to the lack
of universality of Einsteinian LWs.  Regarding the latter observation,
it should be noted that the qualitative behavior of $p(z;N)$ for sums
of independent random variables (or of $p(x,t)$ for LWs) expressed by
eqs.~(\ref{eq6})-(\ref{eq7}) and the corresponding scaling of the
moment hierarchy is universal, in the meaning that they depend
exclusively on the asymptotic scaling exponent $\xi$ controlling
$T(\tau)$ and not on the fine details of this density.

The differential moment exponent spectrum $\nu^*(q)$ provides
a convenient and clear indicator of these two concurring
propagation tendencies. Starting from this compact description
of the higher-order anomalies observed in LWs, a simple approximate
two-phase model has been developed that correctly accounts for the scaling
of the whole moment hierarchy. In principle, the same two-phase
approach can be extended to anomalous diffusive LWs, i.e., in the range
$1 < \xi < 2$, by replacing the Laplacian operator  in eq.~(\ref{eq5_10}) 
with a fractional-order differential operator. This
extension is left for further investigation.

As a final comment, the analogy between CLT and Renormalization Group Theory
envisaged by Jona-Lasinio in 1975 \cite{ren1} suggests that the results of the
present manuscript as regards higher-order anomalies could be usefully transferred
also to the analysis of renormalization methods in quantum field theory.

\appendix 

\section{Further example elucidating the singular limit eq.~(\ref{eq5b})}
\label{sec2bis_1}

We consider the CTRW model of a L\'evy walk, i.e. $x_N=\sum_{h=1}^N r_h \, \tau_h$,
in the case $\tau_h$ are independent of one another (and of the
binary variables $r_h$), but are not identically distributed. 
Each $\tau_h$ is characterized by a density $T_h(\tau)= \xi_h/(1+\tau)^{\xi_h+1}$,
with $\xi_h$ depending on $h$. 
Since 
the variables $\tau_h$ are assumed to possess bounded variances, 
the classical CLT theorem can be  easily extended \cite{clt1}.
In the language of LW this corresponds to a form of aging,
in which the statistical properties of the transition times are functions
of the operational time.

In this case, the kurtosis $\kappa_N$ associated with $x_N$ is given
by
\begin{equation}
\kappa_N= \frac{\sum_{h=1}^N \langle \tau_h^4 \rangle }{\left ( \sum_{h=1}^N \langle \tau_h^2 \rangle  \right )^2} + 3
\label{eqax1}
\end{equation}
where
\begin{equation}
 \hspace{-1cm} \langle \tau_h^2 \rangle= \frac{2}{(\xi_h-1) (\xi_h-2)} \, ,
\qquad
\langle \tau_h^4 \rangle = \frac{24}{(\xi_h-1) (\xi_h -2) (\xi_h-3) (\xi_h-4)}
\label{eqax2}
\end{equation}
Assume that  $\langle \tau^4_h \rangle$ were bounded for any $h=1,2,\dots$, 
while $\lim_{h \rightarrow \infty} \langle \tau^4_h \rangle = \infty$. This corresponds
to the case where $\xi_h>4$, and $\lim_{h \rightarrow \infty} \xi_h=4$. For
instance, one may consider the  situation described by the sequence of exponents
\begin{equation}
\xi_h= 3 + \frac{2}{1+h^\beta}
\label{eqax3}
\end{equation}
\begin{figure}[htbp]
\begin{center}
\resizebox{0.8\columnwidth}{!}{%
 \includegraphics{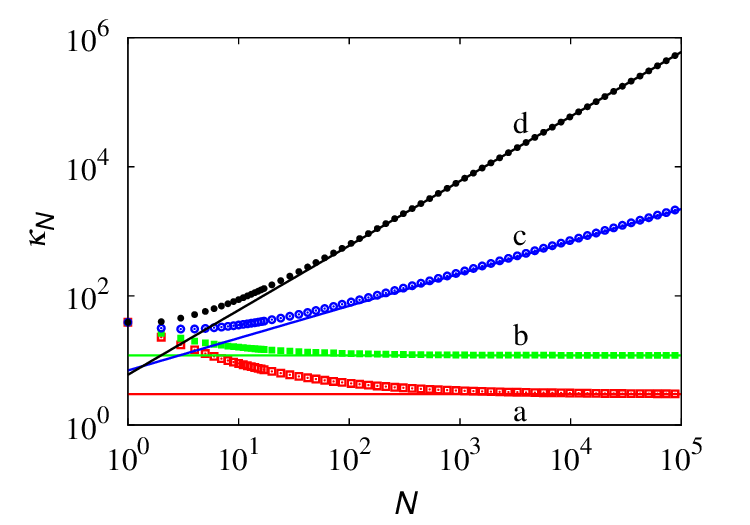}}
\end{center}
\caption{Kurtosis $\kappa_N$ vs $N$ deriving from eq. (\ref{eqax1}), with
$\xi_h$ given by eq. (\ref{eqax3}).
Symbols  (a) refer to $\beta=0.5$, (b) to $\beta=1$,  (c) to $\beta=1.5$,
(d) to $\beta=2$. Solid line (a) corresponds to $\kappa_n=3$, corresponding to the
Gaussian limit, line (b) to $\kappa_n=\mbox{const}\neq 3$, line (c) to
$\kappa_n \sim n^{1/2}$, line (d) to $\kappa_n \sim n$.}
\label{Fig1bis}
\end{figure}

Since $\sum_{h=1}^N  \langle \tau_h^2 \rangle \sim N^2$, and
$\sum_{h=1}^N  \langle \tau_h^4 \rangle \sim \sum_{h=1}^N h^\beta  \sim N^{1+\beta}$,
it follows that
\begin{equation}
\kappa_N  =
\left \{
\begin{array}{lll}
3 \;\; & \;\;  & \beta<1 \\
c \neq 3 & \;\; & \beta=1 \\
d \, N^{\beta-1} +O(N^{\beta-2}) & \;\;& \beta>1
\end{array}
\right .
\label{eqax4}
\end{equation}
where $c,\, d>0$ are constant, and consequently the asymptotic behavior of the fourth-order moment depends towards
the rate of convergence of $\xi_h$ towards the limit value $\xi=4$, determining
unbounded $\langle \tau^4 \rangle$. This phenomenon is depicted in 
figure \ref{Fig1bis}. This  simple analytical example shows the  subtle role
of the singular limit eq. (\ref{eq5b}) controlling the scaling of
the higher-order elements of the moment hierarchy of sums of independent random
variables fulfilling the hypothesis  of the CLT.

 .

\end{document}